\documentclass[aps,prl,showpacs,twocolumn,superscriptaddress]{revtex4}
\usepackage{bm,color}
\usepackage{graphicx}
\usepackage{amsmath}
\usepackage{color}
\pdfoutput=1

\begin{document}
\title {DC spinmotive force from microwave-active resonant dynamics of skyrmion crystal under a tilted magnetic field}
\author{Tatsuya Koide}
\affiliation{Department of Physics and Mathematics, Aoyama Gakuin University, Sagamihara, Kanagawa 229-8558, Japan}
\author{Akihito Takeuchi}
\affiliation{Department of Physics and Mathematics, Aoyama Gakuin University, Sagamihara, Kanagawa 229-8558, Japan}
\author{Masahito Mochizuki}
\affiliation{Department of Applied Physics, Waseda University, Okubo, Shinjuku-ku, Tokyo 169-8555, Japan}
\affiliation{PRESTO, Japan Science and Technology Agency, Kawaguchi, Saitama 332-0012, Japan}
\begin{abstract}
We theoretically show that a temporally oscillating spin-driven electromotive force or voltage with a large DC component can be generated by exciting microwave-active spin-wave modes of a skyrmion crystal confined in a quasi two-dimensional magnet under a tilted magnetic field. The DC component and the AC amplitude of the oscillating electric voltage is significantly enhanced when the microwave frequency is tuned to an eigenfrequency of the peculiar spin-wave modes of the skyrmion crystal called ``rotation modes" and ``breathing mode", whereas the sign of the DC component depends on the microwave polarization and on the spin-wave mode. These results provide an efficient means to convert microwaves to a DC electric voltage by using skyrmion-hosting magnets and to switch the sign of the voltage via tuning the microwave frequency, which are important capabilities for spintronic devices.
\end{abstract}
\pacs{76.50.+g,78.20.Ls,78.20.Bh,78.70.Gq}
\maketitle

\section{Introduction}
\begin{figure}[t]
\begin{center}
\includegraphics[width=1.0\columnwidth]{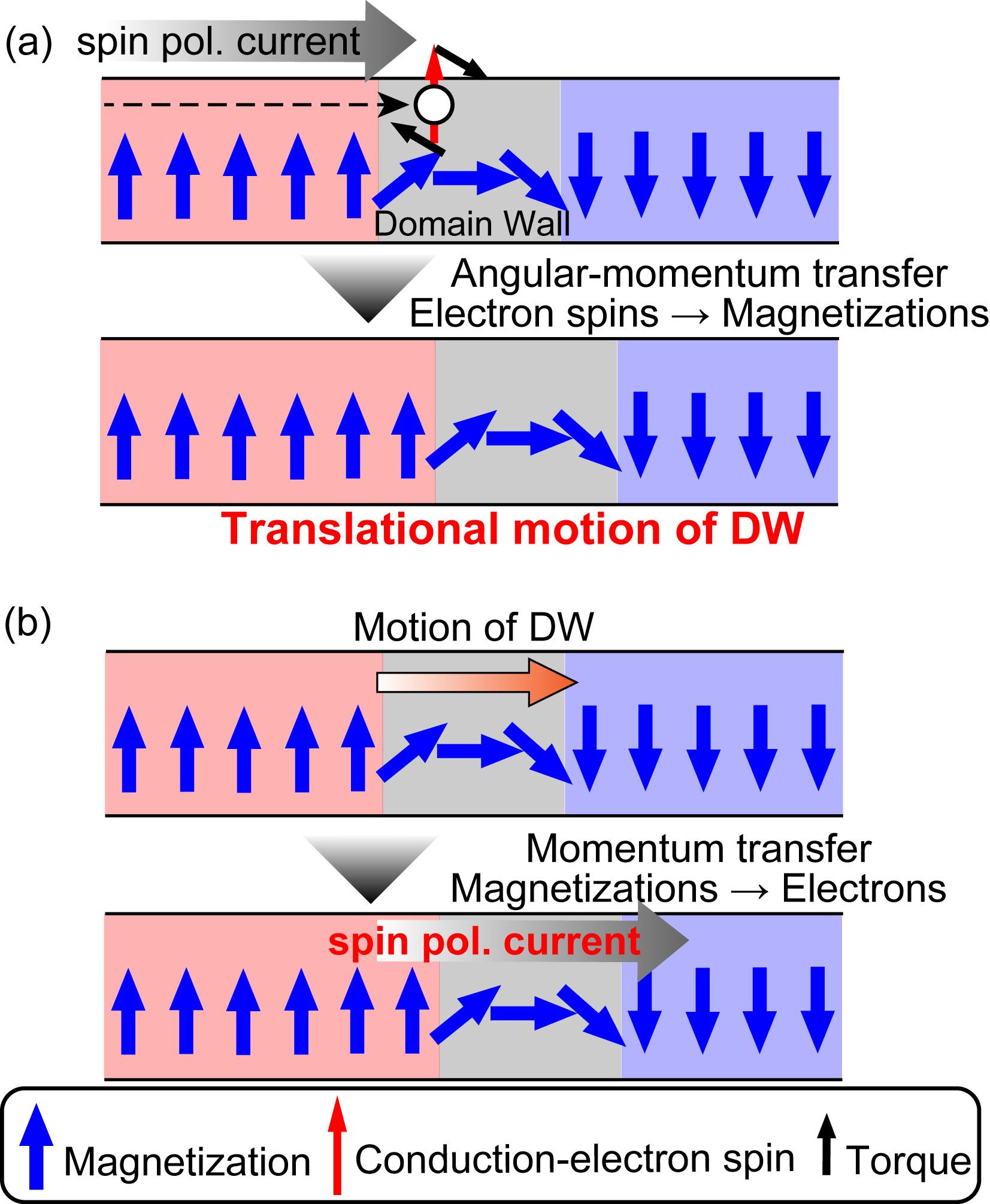}
\caption{(a) Schematics of spin-transfer torque mechanism. Angular-momentum transfer from conduction-electron spins of an injected spin-polarized current to magnetizations constituting a noncollinear magnetic texture drives the translational motion of the magnetic texture. (b) Schematics of spin-driven electromotive force, the so-called ``spinmotive force". Momentum transfer from a driven noncollinear magnetic texture to the conduction electrons via exchange coupling causes an effective electromotive force acting on the conduction electrons, resulting in the generation of spin-polarized electric currents. Here we consider a domain wall in a ferromagnetic nanowire as an example of the noncollinear magnetic texture.}
\label{Fig01}
\end{center}
\end{figure}
Dynamic coupling between conduction electrons and noncollinear magnetic textures in magnets gives rise to a variety of physical phenomena, which have attracted significant research interest from the viewpoints of both fundamental science and technical applications. 
The spin-driven electromotive force (i.e., an emergent electric field induced by magnetization dynamics) is one of the most important of these phenomena~\cite{Volovik87,Barnes07}. It was proposed theoretically that noncollinear magnetizations produces a spatially inhomogeneous effective vector potential acting on conduction electrons via the $s$-$d$ exchange coupling between conduction-electron spins and the magnetization. A temporal variation of this effective vector potential due to the dynamics of the magnetization induces an effective electromotive force that acts on the conduction electrons, which is referred to as the spinmotive force". The spinmotive force can also be interpreted as the inverse effect of the spin-transfer torque mechanism (see Fig.~\ref{Fig01}). The spinmotive force is expressed by
\begin{eqnarray}
E_\mu(\bm r, t)=\frac{\hbar}{2e}\bm m \cdot 
(\partial_\mu \bm m \times \partial _t \bm m)
\quad \quad (\mu=x,y),
\label{eq:smf1}
\end{eqnarray}
where $\bm m(\bm r,t)$ is the normalized classical magnetization vector. This formula explicitly indicates that the magnetization must vary both spatially and temporally to produce the spinmotive force. Several experimental reports discuss the generation and observation of the spinmotive force in ferromagnetic domain walls and magnetic vortices~\cite{SAYang09,Tanabe12}. 

Recent theoretical work revealed that hexagonally packed skyrmions in a skyrmion crystal exhibit peculiar spin-wave modes at microwave frequencies, in which the skyrmions rotate uniformly in a counterclockwise or clockwise sense (the rotation modes), or uniformly expand and shrink in an oscillatory manner (the breathing mode)~\cite{Mochizuki12,Petrova11}. Subsequently, Ohe {\it et al.} showed numerically that the spin-driven electric voltage (the so-called ``spin voltage") can be generated from these microwave-active spin-wave modes of a magnetic skyrmion crystal~\cite{Ohe13,Shimada15}. An advantage of using a skyrmion crystal to generate the spin voltage is that the periodically aligned skyrmions in the skyrmion crystal work as batteries connected in series, which give a large electric voltage by summing each contribution. However, th spin voltage oscillates temporally and is, more specifically, a pure AC voltage with an average of zero. For certain spintronics applications, DC electric voltages are preferable. One possible way to obtain a DC voltage is to use an AC-DC transducer to convert the AC voltage to a DC voltage. However, this approach requires fabricating complicated devices. Moreover, we cannot avoid reduction of the voltage in the conversion process, which can be a critical problem because the spin voltage is originally very small. Therefore, a simple technique to generate a DC spin voltage is highly desired.

In this paper, we show theoretically that an oscillating spin voltage with a large DC component can be generated by exciting the microwave-active spin-wave modes of skyrmion crystal on a quasi-two-dimensional thin-plate magnet under an external magnetic field $\bm H_{\rm ex}$ tilted with respect to the perpendicular direction. By numerically solving the Landau-Lifshitz-Gilbert (LLG) equation, we trace the dynamics of magnetization in the skyrmion crystal activated by a microwave magnetic field $\bm H^\omega$. We use the results of the simulation of the magnetization dynamics to calculate the spatiotemporal profiles of the spinmotive force and the temporal profiles of the spin voltage. The results show that the DC component and the AC amplitude of the temporally oscillating electric voltage increase significantly when the frequency of the microwave is tuned to an eigenfrequenciy of the spin-wave modes, which converts the microwave power to a DC voltage with high efficiency using a skyrmion-hosting magnet. The results also show that the sign of the DC voltage depends on the excited spin-wave modes, which indicates that the sign of the voltage can be swicthed by tuning the microwave frequency. Our finding will be useful for technical applications in spintronics devices.

\section{Model and Method}
We start with a classical Heisenberg model on a square lattice to describe a thin-plate specimen of the skyrmion-hosting magnet. The model contains the ferromagnetic-exchange interaction, the Dzyaloshinskii-Moriya (DM) interaction, and the Zeeman interaction with an external magnetic field $\bm H_{\rm ex}=(H_x, 0, H_z)$~\cite{Bak80}. The Hamiltonian is given by~\cite{YiSD09},
\begin{eqnarray}
\mathcal{H}_0
&=&-J \sum_{i} (\bm m_i \cdot \bm m_{i+\hat{x}}+\bm m_i \cdot \bm m_{i+\hat{y}})\nonumber \\
& &-D \sum_{i} (\bm m_i \times \bm m_{i+\hat{x}} \cdot \hat{\bm x}
+\bm m_i \times \bm m_{i+\hat{y}} \cdot \hat{\bm y})
\nonumber \\
& &-\bm H_{\rm ex} \cdot \sum_i \bm m_i,
\label{eq:model}
\end{eqnarray}
where $\bm m_i$ is the normalized magnetization vector on the $i$th lattice site. We use $J$=1 as the energy units and set $D/J$=0.27. The vertical component of $\bm H_{\rm ex}$ is fixed at $H_z$=0.036, whereas the in-plane component is $H_x=H_z\tan\theta$ with $\theta$ being the tilting angle of $\bm H_{\rm ex}$.

The diameter of a skyrmion is determined by competition between the DM interaction and the ferromagnetic exchange interaction, which favor rotating and parallel magnetization alignments, respectively. A stronger DM interaction causes rapid rotation of the magnetizations and consequently a smaller skyrmion size. Because $\phi\sim D/(\sqrt{2}J)$ holds for the magnetization rotation angle $\phi$, the spatial period in the skyrmion crystal becomes $\lambda_{\rm m}\sim 2\pi a/\phi$ with $a$ being the lattice constant. Therefore, the ratio $D/J=0.27$ gives $\lambda_{\rm m}\sim 18$ nm if we assume a typical lattice constant of $a$=0.5 nm, which corresponds to the experimentally observed skyrmion size in MnSi.

\begin{figure}[t]
\begin{center}
\includegraphics[width=1.0\columnwidth]{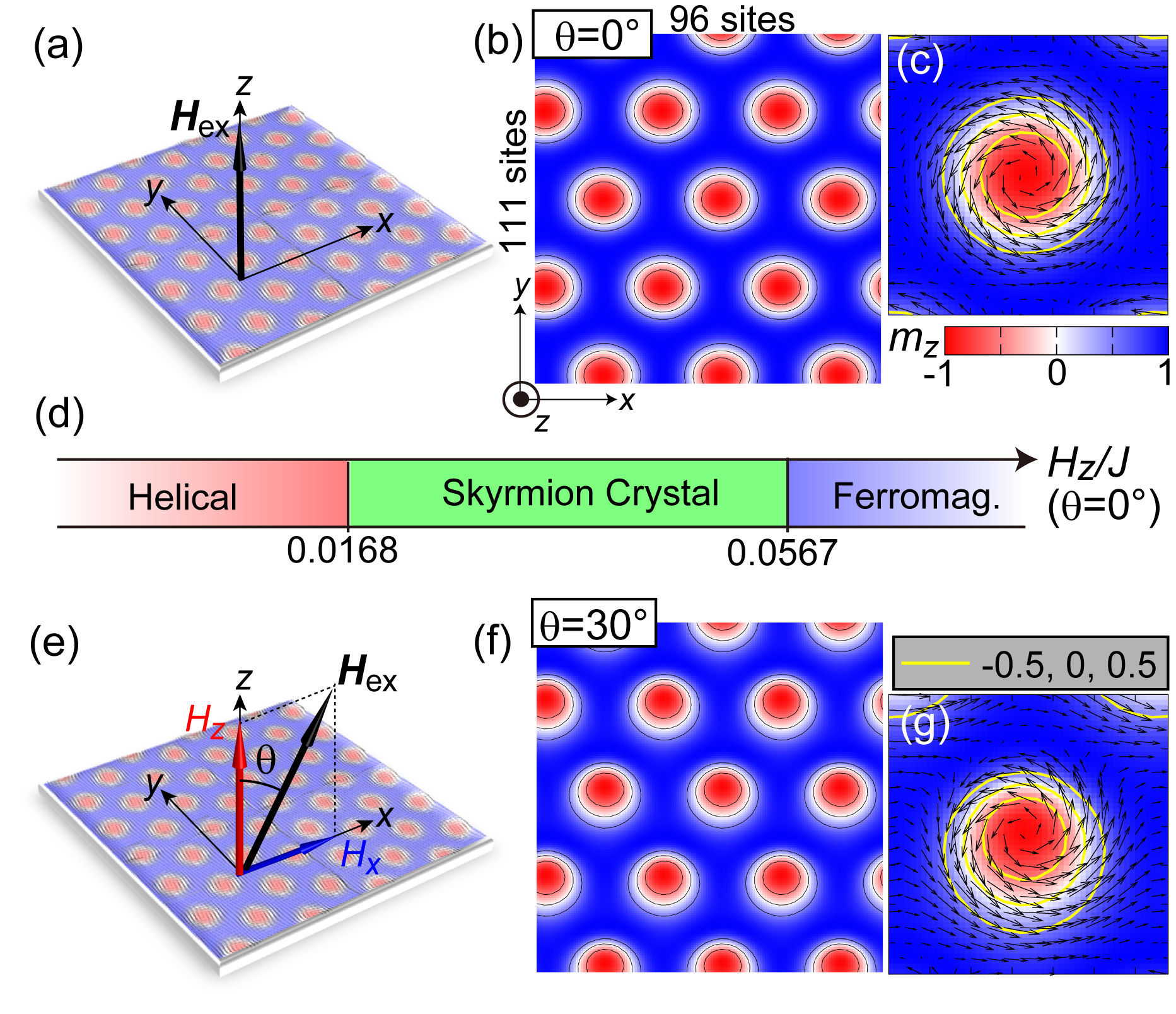}
\caption{(a) [(e)] Thin-plate specimen of chiral-lattice magnet hosting a skyrmion crystal under a perpendicular [tilted] external magnetic field $\bm H_{\rm ex}=(H_z\tan\theta,0,H_z)$ with a tilting angle of $\theta=0^\circ$ [$\theta \ne 0^\circ$]. (b) [(f)] Skyrmion crystal under the perpendicular [tilted] magetic field $\bm H_{\rm ex}$ with $\theta=0^\circ$ [$\theta=30^\circ$]. (c) [(g)] Magnetization configuration of a skyrmion constituting the skyrmion crystal under a perpendicular [tilted] magnetic field $\bm H_{\rm ex}$. (d) Theoretical phase diagram of the spin model given by Eq.~(\ref{eq:model}) at $T=0$ as a function of the out-of-plane component of $\bm H_{\rm ex}$ when $\theta=0^\circ$.}
\label{Fig02}
\end{center}
\end{figure}
When the external magnetic field $\bm H_{\rm ex}$ is applied perpendicular to the thin-plate plane [i.e., $\bm H_{\rm ex}=(0,0,H_z)$] as shown in Fig.~\ref{Fig02}(a), the skyrmion crystal phase with hexagonally packed skyrmions [Fig.~\ref{Fig02}(b)] appears when the field strength $H_z$ is moderate. In this case, each skyrmion in the skyrmion crystal has a circular symmetry as shown in Fig.~\ref{Fig02}(c). Figure~\ref{Fig02}(d) shows a theoretical phase diagram of this spin model given by Eq.~(\ref{eq:model}) at $T$=0 as a function of the magnetic-field strength $H_z$ when $\bm H_{\rm ex}=(0,0,H_z)$. This phase diagram exhibits the skyrmion-crystal phase in a region of moderate field strength sandwiched by the helical phase and the field-polarized ferromagnetic phase.
The unit conversions when $J$=1 meV are summarized in Table~\ref{tab:uconv}.
\begin{table}
\caption{Unit conversion table when $J$=1 meV.}
\begin{tabular}{l|cc} \hline \hline
               & Dimensionless & Corresponding value \\
               & quantity & with units \\
\hline
Exchange int.  & \hspace{0.5cm} $J$=1  & \hspace{0.5cm} $J$=1 meV \\
Time           & \hspace{0.5cm} $t$=1  & \hspace{0.5cm} $\hbar/J$=0.66 ps \\
Frequency $f=\omega/2\pi$ &\hspace{0.5cm} 
$\omega$=1 & \hspace{0.5cm} $J/h$=241 GHz\\
Magnetic field & \hspace{0.5cm} $H$=1 & \hspace{0.5cm} $J/g\mu_{\rm B}$=8.64 T \\ 
\hline \hline
\end{tabular}
\label{tab:uconv}
\end{table}
When $J$=1 meV, $H_z=1$ corresponds to $\sim$8.64 T. Therefore, the threshold fields of $H_{\rm c1}$=0.0168 and $H_{\rm c2}$=0.0567 in this theoretical phase diagram correspond to 0.145 T and 0.49 T, respectively. These values coincide well with the experimentally observed threshold fields of $\sim$0.15 T and $\sim$0.45 T for MnSi at low temperatures.

The skyrmion crystal phase survives even when the magnetic field $\bm H_{\rm ex}$ is tilted with respect to the perpendicular direction [see Fig.~\ref{Fig02}(e)]. Figure~\ref{Fig02}(f) shows a skyrmion crystal under a magnetic field $\bm H_{\rm ex}$ tilted towards the $x$ direction at a tilting angle of $\theta=30^\circ$, and Fig.~\ref{Fig02}(g) displays the magnetization configuration of a skyrmion in this skyrmion crystal, which has a disproportionate weight in the distribution of the $m_z$ components slanted from its center.

We simulate the dynamics of the magnetization in the skyrmion crystal state activated by a microwave magnetic field $\bm H^\omega$ by numerically solving the LLG equation using the fourth-order Runge-Kutta method. The LLG equation
\begin{equation}
\frac{d\bm m_i}{dt}=-\gamma \bm m_i \times \bm H^{\rm eff}_i 
+\frac{\alpha_{\rm G}}{m} \bm m_i \times \frac{d\bm m_i}{dt}.
\label{eq:LLGEQ}
\end{equation} 
The first term of the right-hand side is the gyrotropic term and describes the precessional motion of magnetizations $\bm m_i$ around the effective local magnetic field $\bm H^{\rm eff}_i$.
The second term is the Gilbert-damping term introduced phenomenologically to describe the dissipation of the gyration energy. The Gilbert-damping coefficient is fixed at $\alpha_{\rm G}=0.02$ for calculations of the microwave absorption spectra shown in Fig.~\ref{Fig03} to precisely evaluate the eigenfrequencies of the skyrmion spin-wave modes, while it is fixed at $\alpha_{\rm G}=0.04$ for calculations of the spinmotive force to evaluate its realistic values for MnSi. The effective magnetic field $\bm H_i^{\rm eff}$ is calculated from the Hamiltonian $\mathcal{H}$=$\mathcal{H}_0$+$\mathcal{H}'(t)$ as follows:
\begin{equation}
\bm H^{\rm eff}_i = -\frac{1}{\gamma \hbar}
\frac{\partial \mathcal{H}}{\partial \bm m_i}.
\label{eq:EFFMF}
\end{equation}
Here $\mathcal{H}_0$ is the model Hamiltonian given by Eq.~(\ref{eq:model}), whereas $\mathcal{H}'(t)$ is the coupling between magnetizations and a time-dependent magnetic field or a microwave magnetic field $\bm H(t)$ in the form,
\begin{equation}
\mathcal{H}'(t)=-\bm H(t) \cdot \sum_i \bm m_i.
\label{eq:TDEPH}
\end{equation}

Using the calculated spatiotemporal dynamics of excited magnetization $\bm m_i$, we calculate spatiotemporal profile of the spinmotive force $\bm E$. For the numerical calculation, it is convenient to rewrite Eq.~(\ref{eq:smf1}) in discretized form as
\begin{align}
&E_{\mu,i}(t)=\frac{\hbar}{2e}\bm m_i(t)
\nonumber \\
&\cdot \left(\frac{\bm m_{i+\hat{\mu}}(t)-\bm m_{i-\hat{\mu}}(t)}{2a}
\times
\frac{\bm m_i(t+\Delta t)-\bm m_i(t-\Delta t)}{2\Delta t}
\right),
\label{eq:smf2}
\end{align}
where $\mu$=$x,y$ and $a$(=5 \AA) is the lattice constant. We also calculate time profiles of the spin voltage by numerically solving the Poisson equation. All the calculations are done using a system of $N=96 \times 111$ sites with periodic boundary conditions.

To identify the spin-wave modes and their eigenfrequencies, we first calculate the dynamical magnetic susceptibility,
\begin{eqnarray}
\chi_\mu(\omega) =
\frac{\Delta M_\mu(\omega)}{H_\mu(\omega)} \quad (\mu=x, y, z),
\end{eqnarray}
where $H_\mu(\omega)$ and $\Delta M_\mu(\omega)$ are the Fourier transforms of the time-dependent magnetic field $\bm H(t)$ and the simulated time-profile of the net magnetization $\Delta \bm M(t)=\bm M(t)-\bm M(0)$ with $\bm M(t)=\frac{1}{N}\sum_{i=1}^N \bm m_i(t)$. In this calculation, we use a short rectangular pulse for $\bm H(t)$ whose components are given by,
\begin{eqnarray}
H_\mu(t)=\left\{
\begin{aligned}
& H_{\rm pulse} \quad & 0 \le t \le 1 \\
& 0 \quad & {\rm others}
\end{aligned}
\right.
\end{eqnarray}
where $t=(J/\hbar)\tau$ is the dimensionless time with $\tau$ being the real time. An advantage of using a short pulse is that the Fourier component $H_\mu(\omega)$ becomes constant being independent of $\omega$ up to first order in $\omega \Delta t$ for a sufficiently short duration $\Delta t$ with $\omega \Delta t \ll 1$. The Fourier component is
\begin{eqnarray}
H_\mu(\omega)&=&\int_0^{\Delta t}\;H_{\rm pulse} e^{i\omega t}dt
=\frac{H_{\rm pulse}}{i\omega}\left(e^{i\omega \Delta t}-1 \right)
\nonumber \\
&\sim&H_{\rm pulse} \Delta t.
\end{eqnarray}
Consequently, we obtain the relationship $\chi_\mu(\omega) \propto \Delta M_\mu(\omega)$.

\section{Results}
\begin{figure}
\begin{center}
\includegraphics[width=1.0\columnwidth]{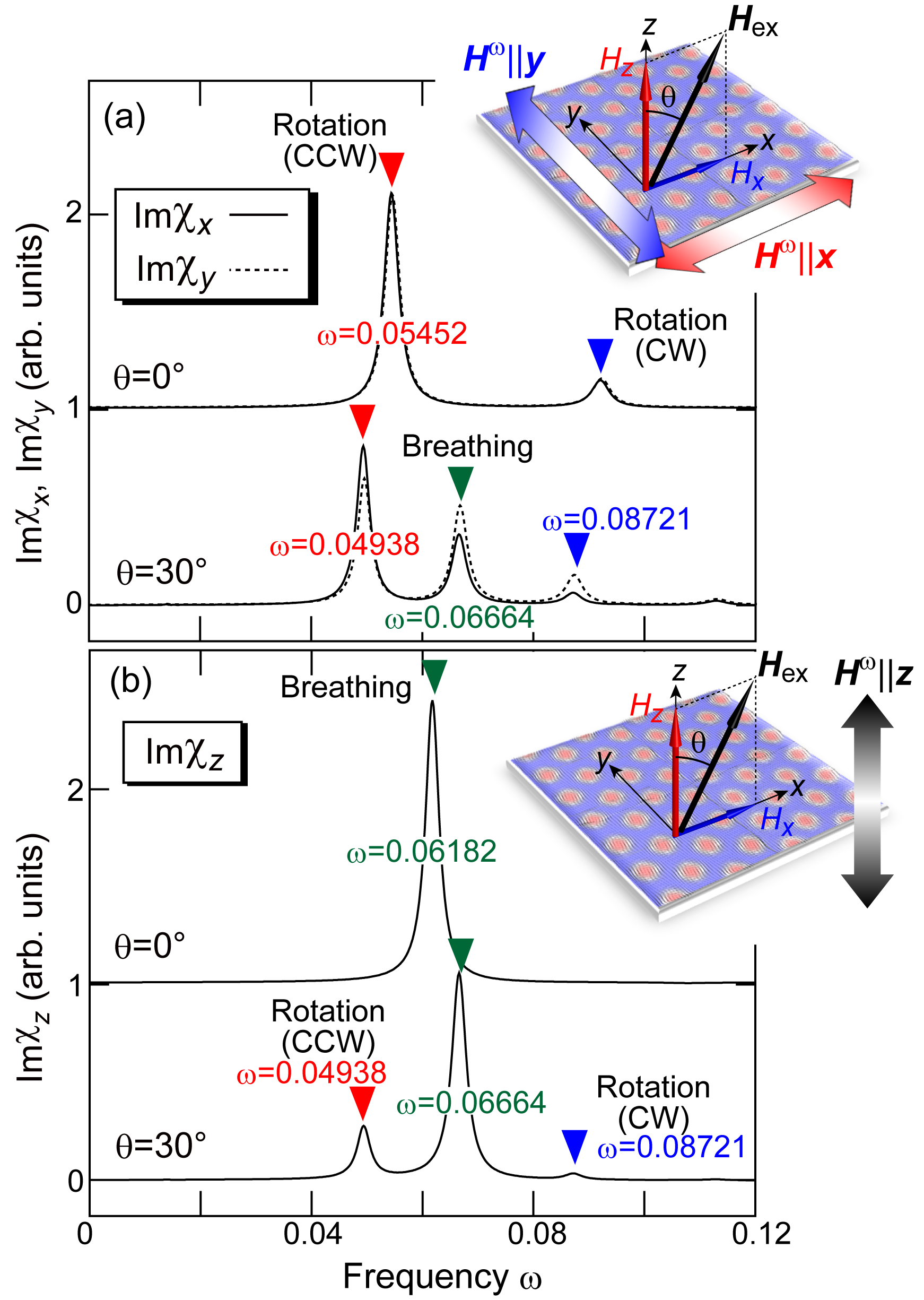}
\caption{Calculated imaginary parts of the dynamical magnetic susceptibilities Im$\chi_\mu$ ($\mu$=$x$,$y$,$z$) of the skyrmion crystal confined in the two-dimensional system under perpendicular ($\theta$=0$^{\circ}$) and tilted ($\theta$=30$^{\circ}$) magnetic fields $\bm H_{\rm ex}$ with $H_z$=0.036 as functions of the microwave frequency $\omega$. (a) Im$\chi_x$ and Im$\chi_y$ for the in-plane microwave polarization with $\bm H^\omega$$\parallel$$\bm x,\bm y$. (b) Im$\chi_z$ for the out-of-plane microwave polarization with $\bm H^\omega$$\parallel$$\bm z$. The dominant mode component and the value of the eigenfrequency are shown at each peak position.}
\label{Fig03}
\end{center}
\end{figure}
\begin{figure*}
\begin{center}
\includegraphics[scale=0.5]{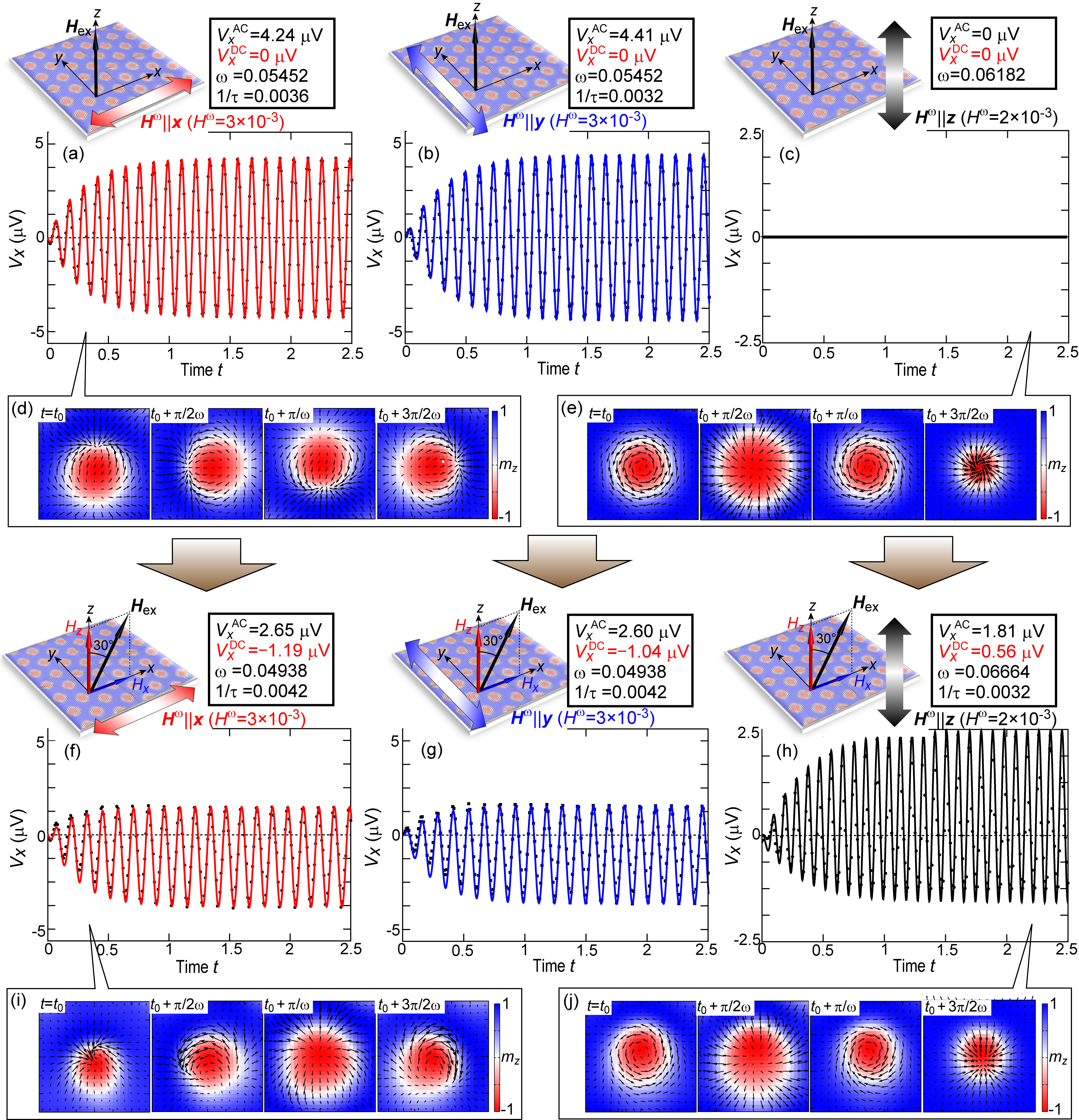}
\caption{(a)-(c) [(f)-(h)] Calculated time profiles of spin voltages induced by the microwave-active spin-wave modes of a skyrmion crystal confined in a thin-plate magnet under a perpendicular [tilted] magnetic field $\bm H_{\rm ex}$ with $\theta=0^\circ$ [$\theta=30^\circ$] and $H_z$=0.036 for various microwave polarizations and frequencies. (a) [(f)] $\bm H^\omega$$\parallel$$\bm x$ with a frequency fixed at the eigenfrequency of the counterclockwise rotation mode, (b) [(g)] $\bm H^\omega$$\parallel$$\bm y$ with a frequency fixed at the eigenfrequency of the counterclockwise rotation mode, and (c) [(h)] $\bm H^\omega$$\parallel$$\bm z$ with a frequency fixed at the eigenfrequency of the breathing mode. The amplitude of the applied microwave is fixed at $H^\omega$=$3\times10^{-3}$ for the in-plane polarized microwave $\bm H^\omega$$\parallel$$\bm x,\bm y$ in panels (a), (b), (f), and (g), whereas it is fixed at $H^\omega$=$2\times10^{-3}$ for the out-of-plane polarized microwave $\bm H^\omega$$\parallel$$\bm z$ in panels (c) and (h). Here, the values of the AC amplitude $V_x^{\rm AC}$, the DC component $V_x^{\rm DC}$, the angular frequency $\omega(=2\pi f)$, and the decay rate $\tau$ of the generated time-dependent spin voltage are shown for each case, as obtained by fitting the simulation results (see text). The results show that the DC voltage $V_x^{\rm DC}$ is always zero under the perpendicular magnetic field $\bm H_{\rm ex}$, whereas it becomes finite when the magnetic field $\bm H_{\rm ex}$ is tilted. (d), (e) [(f), (i)] Snapshots of the spatial distribution of the spinmotive force (arrows) at selected times. The out-of-plane magnetization components $m_z$ are also shown in color to facilitate the visualization of the temporally varying skyrmion shape.}
\label{Fig04}
\end{center}
\end{figure*}
\begin{figure*}
\begin{center}
\includegraphics[scale=0.5]{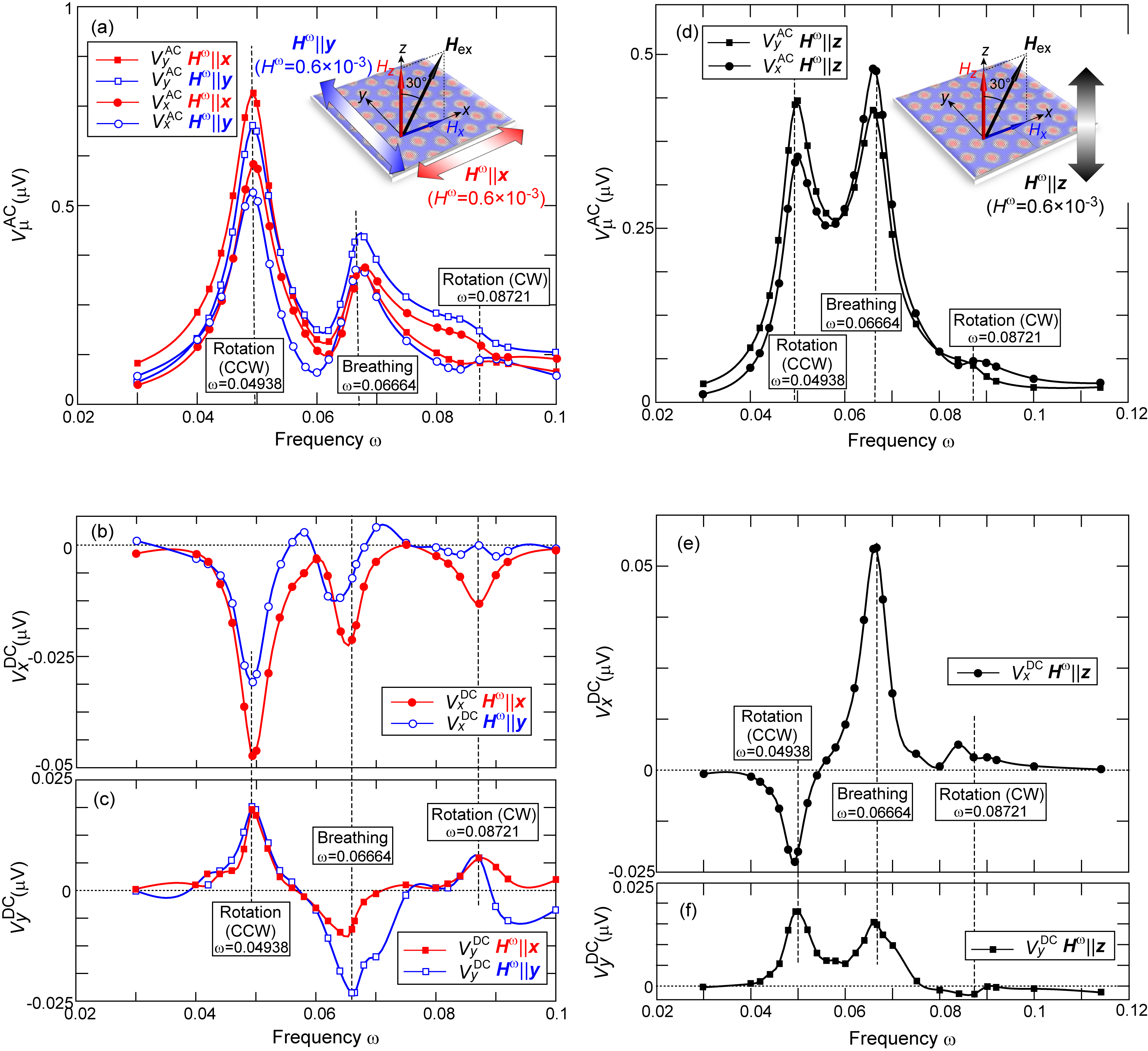}
\caption{(a)-(c) [(d)-(f)] Calculated microwave-frequency dependence of the AC amplitude $V_\mu^{\rm AC}$ and the DC component $V_\mu^{\rm DC}$ ($\mu$=$x$,$y$) of the spin voltage under application of a tilted magnetic field $\bm H_{\rm ex}$ with $\theta$=30$^\circ$ and $H_z$=0.036 for the in-plane [out-of-plane] microwave polarizations $\bm H^\omega$$\parallel$$\bm x,\bm y$ [$\bm H^\omega$$\parallel$$\bm z$], i.e., (a) [(d)] $V_x^{\rm AC}$ and $V_y^{\rm AC}$, (b) [(e)] $V_x^{\rm DC}$, and (c) [(f)] $V_y^{\rm DC}$. Here the amplitude of the applied microwave is fixed at $H^\omega$=$0.6\times10^{-3}$ for all simulations.}
\label{Fig05}
\end{center}
\end{figure*}
Figures~\ref{Fig03}(a) and \ref{Fig03}(b) show the calculated microwave absorption spectra (i.e., the imaginary parts of the dynamical magnetic susceptibilities Im $\chi_\mu$ ($\mu$=$x,y,z$) under perpendicular ($\theta$=0$^{\circ}$) and tilted ($\theta$=30$^{\circ}$) magnetic fields $\bm H_{\rm ex}$ as functions of microwave frequency $\omega(=2\pi f)$. More specifically, Fig.~\ref{Fig03}(a) shows the calculated microwave absorption spectra for the in-plane microwave polarization with $\bm H^\omega$$\parallel$$\bm x,\bm y$ (Im$\chi_x$ and Im$\chi_y$), whereas Fig.~\ref{Fig03}(b) shows the calculated microwave absorption spectra for the out-of-plane microwave polarization with $\bm H^\omega$$\parallel$$\bm z$ (Im$\chi_z$). Here we fix $H_z$=0.036 for the calculations. A dominant component of the oscillation mode is indicated at each peak position. Figure~\ref{Fig03}(a) shows that whereas only two rotation modes can be excited by the in-plane microwave field $\bm H^\omega$$\parallel$$\bm x,\bm y$ under a perpendicular magnetic field $\bm H_{\rm ex}$ ($\theta$=0$^{\circ}$), the breathing mode also becomes active when the magnetic field $\bm H_{\rm ex}$ is tilted. Similarly, Fig.~\ref{Fig03}(b) indicates that the rotation modes can be excited by the out-of-plane microwave field $\bm H^\omega$$\parallel$$\bm z$ as well under the tilted magnetic field $\bm H_{\rm ex}$, although only the breathing mode is active for $\bm H^\omega$$\parallel$$\bm z$ under the perpendicular magnetic field $\bm H_{\rm ex}$.

Comparing the spectra in Fig.~\ref{Fig03}(a) with those in Fig.~\ref{Fig03}(b), we find that three spin-wave modes excited by $\bm H^\omega$$\parallel$$\bm x,\bm y$ have identical eigenfrequencies with three corresponding modes excited by $\bm H^\omega$$\parallel$$\bm z$, indicating that both the in-plane microwave fields $\bm H^\omega$$\parallel$$\bm x,\bm y$ and the out-of-plane microwave field $\bm H^\omega$$\parallel$$\bm z$ excite the same modes under the tilted magnetic field $\bm H_{\rm ex}$. Although the spectra here are calculated for the Bloch-type skyrmion crystal, the Neel-type and the antivortex-type skyrmion crystals produce
 the same spectra.

Figures~\ref{Fig04}(a)-\ref{Fig04}(c) show time profiles of the spin voltage $V_x$ measured between the left and right edges of the system along the $x$ axis under application of a perpendicular magnetic field $\bm H_{\rm ex}$ with $\theta$=0$^\circ$ for various spin-wave modes, i.e., (a) the counterclockwise rotation mode excited by $\bm H^\omega$$\parallel$$\bm x$, (b) the counterclockwise rotation mode excited by $\bm H^\omega$$\parallel$$\bm y$, and (c) the breathing mode excited by $\bm H^\omega$$\parallel$$\bm z$. The dots are the results of the numerical simulations, and the solid lines are fits using the following formula of a forced oscillation with damping:
\begin{eqnarray}
V_\mu=V_\mu^{\rm DC}+V_\mu^{\rm AC}(1-e^{-t/\tau})\sin\omega t
\label{eq:FO}
\end{eqnarray}
with $\mu=x$. Here $V_\mu^{\rm DC}$, $V_\mu^{\rm AC}$, $\omega(=2\pi f)$, and $\tau$ are the DC component, the AC amplitude, the angular frequency, and the decay rate of the induced temporally oscillating spin voltage, respectively. Here, the simulations were done for the microwave amplitude of $H^\omega=3 \times 10^{-3}$ [$H^\omega=2 \times 10^{-3}$] for the in-plane [out-of-plane] polarized microwaves $\bm H^\omega$$\parallel$$\bm x,\bm y$ [$\bm H^\omega$$\parallel$$\bm z$], which corresponds to $\sim$26 mT [$\sim$17.3 mT] when $J$=1 meV. When the magnetic field $\bm H_{\rm ex}$ is perpendicular to the thin-plate-shaped sample, pure AC voltages with zero DC component are obtained for the rotation mode as seen in Figs.~\ref{Fig04}(a) and \ref{Fig04}(b), whereas the spin voltage is exactly zero for the breathing mode as seen in Fig.~\ref{Fig04}(c). For example, the data in Fig.~\ref{Fig04}(a) are well fit by Eq.~(\ref{eq:FO}) with $V_x^{\rm DC}=0$ $\mu$V and $V_x^{\rm AC}=4.24$ $\mu$V.

In Figs.~\ref{Fig04}(d) and \ref{Fig04}(e), spatial maps of the electromotive force vectors $\bm E(\bm r, t)$ at selected moments are shown by arrows. Here the colors indicate the out-of-plane magnetization component $m_z$ to visualize the temporally varing skyrmion shape at each moment in the spin-wave excitation. Figure~\ref{Fig04}(d) shows that the $\bm E$-field vectors circulate together with the rotating skyrmion in the counterclockwise rotation mode. The observed pure AC voltage is attributed to this symmetric circulation of the $\bm E$-field vectors. On the contrary, Fig.~\ref{Fig04}(e) shows that the circularly symmetric source and sink of the $\bm E$ fields oscillate together with the oscillatory expansion and shrinkage of the skyrmion in the breathing mode. The observed exact-zero electric voltage is attributed to the radially distributed $\bm E$-field vectors with a circular symmetry.

Conversely, Figs.~\ref{Fig04}(f)-\ref{Fig04}(h) show the temporal profiles of spin voltage $V_x$ under application of a tilted magnetic field $\bm H_{\rm ex}$ with $\theta$=30$^\circ$. Figures~\ref{Fig04}(f) and \ref{Fig04}(g) show that the center of mass of the oscillation of $V_x$ shifts downwards indicating the emergence of a finite DC component $V_x^{\rm DC}$. Surprisingly, the oscillating electric voltage with a large DC component also appears for the breathing mode as seen in Fig.~\ref{Fig04}(h), which is in striking contrast with the pure AC voltage with zero DC component observed under the perpendicular magnetic field $\bm H^{\rm ex}$. Note that the DC component $V_x^{\rm DC}$ is negative in the former cases in Figs.~\ref{Fig04}(f) and \ref{Fig04}(g), whereas it is positive in the latter case in Fig.~\ref{Fig04}(h).

Figures~\ref{Fig05}(a) and \ref{Fig05}(b) show the calculated microwave-frequency dependence of the AC-amplitude of the spin voltage $V_\mu^{\rm AC}$ ($\mu$=$x$,$y$) for the in-plane microwave polarizations $\bm H^\omega$$\parallel$$\bm x,\bm y$ and the out-of-plane microwave polarization $\bm H^\omega$$\parallel$$\bm z$, respectively. We fixed the amplitude of the applied microwave field at $H^\omega=0.6\times 10^{-3}$ in the simulations. These plots indicate that the AC amplitude is enhanced and peaks at eigenfrequencies of the spin-wave modes. In particular, significant enhancement occurs when the microwave frequency is tuned to the eigenfrequency of the counterclockwise rotation mode and that of the breathing mode.

Conversely, Figs.~\ref{Fig05}(c)-\ref{Fig05}(f) show the microwave-frequency dependence of the DC components of the spin voltage $V_\mu^{\rm DC}$ ($\mu$=$x$,$y$) for different microwave polarizations. Specifically, Fig.~\ref{Fig05}(c) shows $V_x^{\rm DC}$ for $\bm H^\omega$$\parallel$$\bm x,\bm y$, Fig.~\ref{Fig05}(d) shows $V_y^{\rm DC}$ for $\bm H^\omega$$\parallel$$\bm x,\bm y$, Fig.~\ref{Fig05}(e) shows $V_x^{\rm DC}$ for $\bm H^\omega$$\parallel$$\bm z$, and Fig.~\ref{Fig05}(f) shows $V_y^{\rm DC}$ for $\bm H^\omega$$\parallel$$\bm z$. The microwave amplitude was again fixed at $H^\omega=0.6\times 10^{-3}$ for these simulations. The results show that the DC component is enhanced at the eigenfrequencies of the spin-wave modes. Moreover, the sign of $V_\mu^{\rm DC}$ depends on the spin-wave mode. For example, the sign variation is apparent for the three spin-wave modes in the profile of $V_y^{\rm DC}$ for $\bm H^\omega$$\parallel$$\bm x,\bm y$ [Fig.~\ref{Fig05}(d)] and in the profile of $V_x^{\rm DC}$ for $\bm H^\omega$$\parallel$$\bm z$ [Fig.~\ref{Fig05}(e)], which indicates that the sign of the voltage depends strongly on the mode. 

The results also show that the sign of $V_\mu^{\rm DC}$ depends on the microwave polarization, even when the same spin-wave mode is excited. For example, for the breathing mode, both $V_x^{\rm DC}$ and $V_y^{\rm DC}$ are negative for the in-plane microwave polarization $\bm H^\omega$$\parallel$$\bm x,\bm y$ as seen in Figs.~\ref{Fig05}(c) and \ref{Fig05}(d), but are positive for the out-of-plane microwave polarization $\bm H^\omega$$\parallel$$\bm z$ as seen in Figs.~\ref{Fig05}(e) and \ref{Fig05}(f). Finally, note that a large DC voltage is obtained for the counterclockwise rotation mode excited by $\bm H^\omega$$\parallel$$\bm x,\bm y$ and for the breathing mode excited by $\bm H^\omega$$\parallel$$\bm z$, which indicates that these sets of microwave polarization and the spin-wave mode are suitable for efficient conversion of microwaves to a DC electric voltage.

\begin{figure}
\begin{center}
\includegraphics[width=1.0\columnwidth]{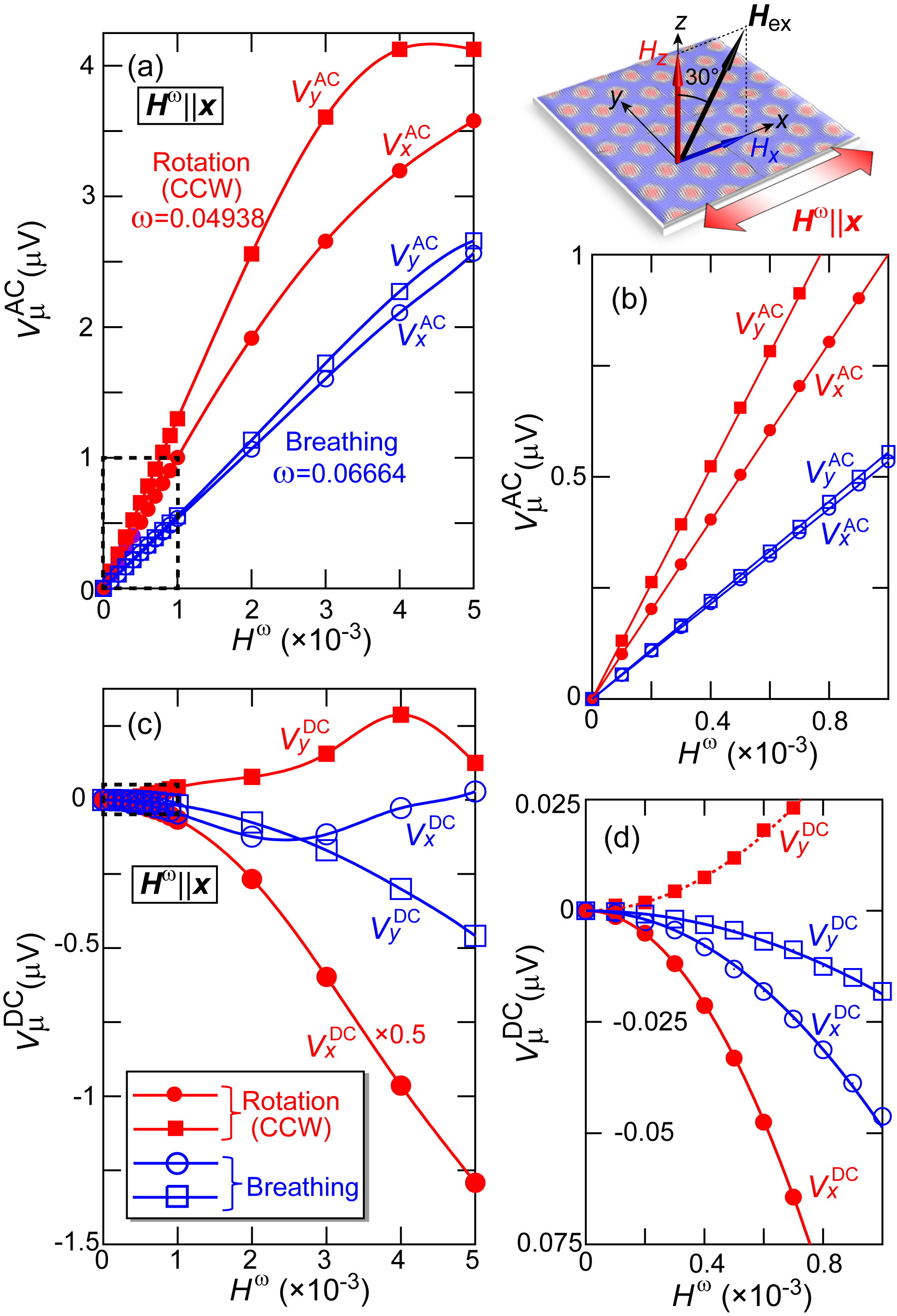}
\caption{(a) [(c)] Calculated microwave-amplitude dependence of the AC amplitudes [the DC components] of the spin voltages $V_x^{\rm AC}$ and $V_y^{\rm AC}$ [$V_x^{\rm DC}$ and $V_y^{\rm DC}$] for the counterclockwise rotation mode and the breathing mode excited by the in-plane microwave field $\bm H^\omega$$\parallel$$\bm x$ under application of a tilted magnetic field $\bm H_{\rm ex}$ with $\theta$=30$^\circ$ and $H_z$=0.036. (b) [(d)] Magnified view of the area indicated by the dashed rectangle in panel (a) [panel (c)]. The frequency of the applied microwave field is fixed at $\omega$=0.04938 for the counterclockwise rotation mode, and at $\omega$=0.0664 for the breathing mode.}
\label{Fig06}
\end{center}
\end{figure}
Figure~\ref{Fig06}(a) plots the calculated AC amplitudes of spin voltages $V_x^{\rm AC}$ and $V_y^{\rm AC}$ as functions of the microwave amplitude $H^\omega$ for two different spin-wave modes (i.e., the counterclockwise rotation mode and the breathing mode for the in-plane microwave polarization $\bm H^\omega$$\parallel$$\bm x$). Figure~\ref{Fig06}(b) provides an expanded view of the area for small $H^\omega$ indicated by the dashed rectangle in Fig.~\ref{Fig06}(a). The frequency of the applied microwave field is fixed at $\omega$=0.04938 for the counterclockwise rotation mode, whereas at $\omega$=0.0664 for the breathing mode in the simulations. When the microwave field is weak with small $H^\omega$, the AC amplitude $V_\mu^{\rm AC}$ is proportional to $H^\omega$. Deviations from the linear relation appear when $H^\omega$ becomes large as seen above $H^\omega \sim 3\times10^{-3}$ in Fig.~\ref{Fig06}(a). This deviation can be attributed to the distortion of the triangular skyrmion crystal due to the intense spin-wave excitations. 

Conversely, Fig.~\ref{Fig06}(c) shows the DC components of spin voltages $V_x^{\rm DC}$ and $V_y^{\rm DC}$ as a function of $H^\omega$ for the two different spin-wave modes. Figure~\ref{Fig06}(d) again expands the area of small $H^\omega$ indicated by the dashed rectangle in Fig.~\ref{Fig06}(c). Figures~\ref{Fig06}(c) and \ref{Fig06}(d) show that the DC voltages $V_\mu^{\rm DC}$ are proportional to the square of $H^\omega$ when $H^\omega$ is small, whereas a deviation from this scaling law appears again for a strong microwave field with a large $H^\omega$ because the triangular array of the skyrmion crystal becomes distorted by the intense spin-wave excitations.

\begin{figure}
\begin{center}
\includegraphics[width=1.0\columnwidth]{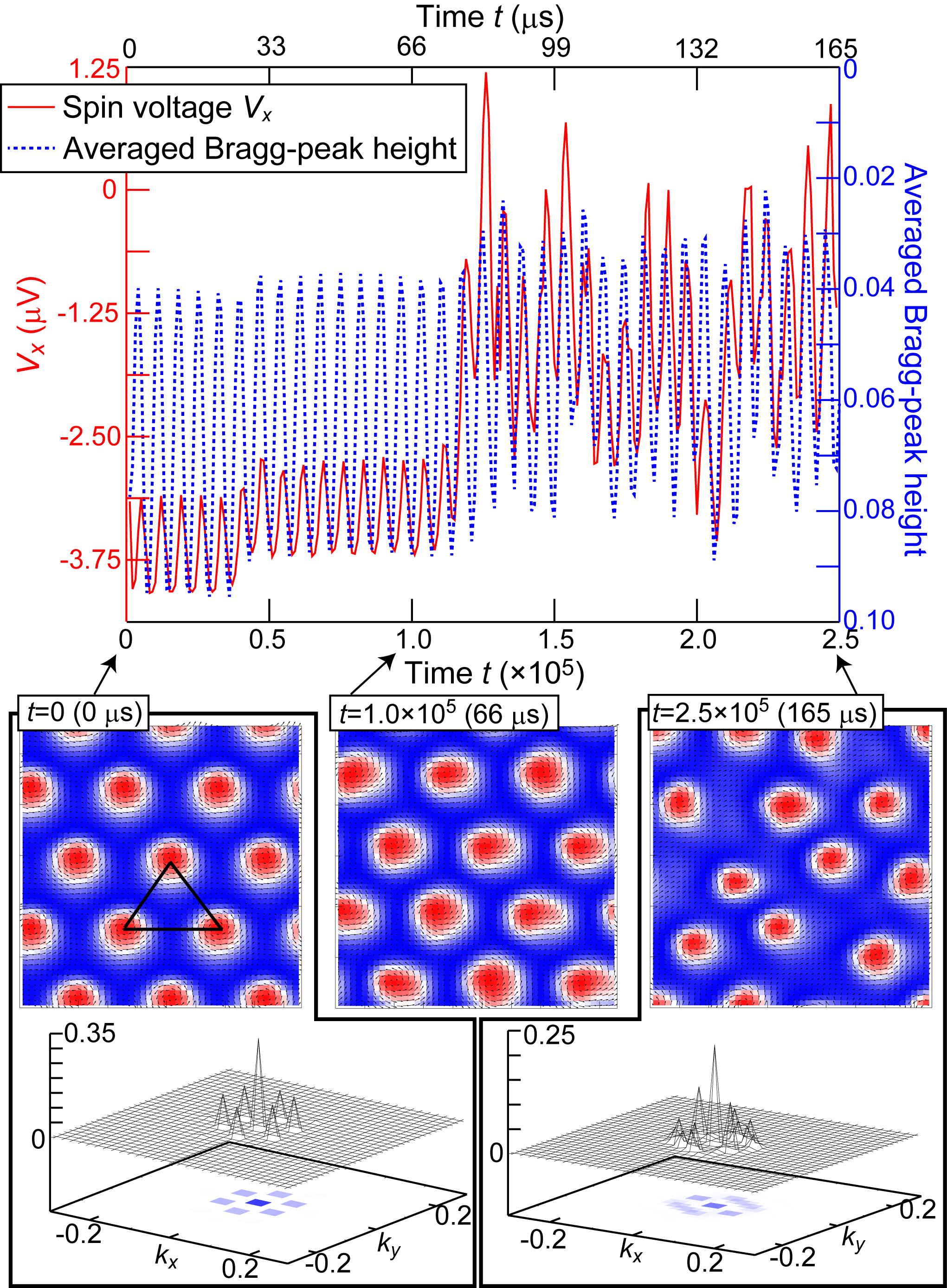}
\caption{Time profiles of the spin voltage and the average of six Bragg peaks. Snapshots of the skyrmion-crystal configurations and the Bragg peaks in the momentum space at selected moments are also displayed. In the simulations, the skyrmion crystal is continuously excited by an intense in-plane polarized microwave $\bm H^\omega$$\parallel$$\bm x$ with $H^\omega=3\times 10^{-3}$ and $\omega$=0.04938 under application of a tilted magnetic field $\bm H_{\rm ex}$ with $\theta$=30$^\circ$ and $H_z$=0.036.}
\label{Fig07}
\end{center}
\end{figure}
Finally we propose that measurements of the spin voltage may provide a powerful tool to detect several transition phenomena of magnetic skyrmions. It is known that skymions exhibit rich structural phase transitions. For example, nonequilibrium triangular-square lattice structural transitions upon rapid cooling have been observed experimentally~\cite{Karube16,Nakajima17,Oike16,Chacon18}. Other issues of interest are the melting of a skyrmion crystal due to intense spin-wave excitations~\cite{Mochizuki12} and defect formation in a skyrmion crystal during the crystallization~\cite{Matsumoto16,Nakajima17b} are also issues of interest. We demonstrate that the time profile of spin voltage sensitively varies depending on the spatial configuration of skyrmions in the skyrmion crystal. In Fig.~\ref{Fig07}, we show a simulated time profile of the spin voltage and that of the average of six Bragg peaks together. Snapshots of the skyrmion-crystal configuration and the Bragg peaks at selected moments are also shown. In the simulations, the skyrmion crystal is continuously excited by an intense in-plane polarized microwave $\bm H^\omega$$\parallel$$\bm x$ with $H^\omega=3\times 10^{-3}$ and $\omega$=0.04938 under application of a tilted magnetic field $\bm H_{\rm ex}$ with $\theta$=30$^\circ$. With this intense microwave field, the spin-wave amplitude or the circulation radius of skyrmions in the resonantly excited rotation mode exceeds the lattice constant of the skyrmion crystal, which eventually results in the distortion of the triangular skyrmion crystal. We find an apparent correlation between the oscillation fashion of the spin voltage and that of the Bragg peak in their temporal profiles. Although the skyrmion crystal maintains its triangular form for a while after the onset of microwave irradiation, both the spin voltage and the average magnitude of the Bragg peaks oscillate in a steady fashion. On the contrary, when the skyrmion crystal becomes distorted after a sufficient duration of the microwave application, both quantities undergo disordered oscillations. Although further studies are required to determined how to detect the randomness or the degree of disorder of skyrmion crystals via spin-voltage measurements, this technique may spawn a new research field involving nontrivial phase-transition phenomena of magnetic skyrmions, such as jamming transitions, liquid-gas transitions, dynamical structural transitions and so on, which are issues of future interest.

\section{Summary}
To summarize, we theoretically showed that a temporally oscillating spinmotive force and spin voltage with a large DC component can be generated by exciting spin-wave modes of a magnetic skyrmion crystal confined in a two-dimensional magnet under a static magnetic field tilted from the perpendicular direction. The DC component and the AC amplitude of the induced electric voltage are significantly enhanced, when the frequency of the applied microwave is tuned to an eigenfrequency of the spin-wave modes. 

The results also revealed that the sign of the DC voltage depends on the microwave polarization and on the excited spin-wave mode, which makes it possible to switch the voltage sign by tuning the microwave parameters. Crucially, the periodically aligned skyrmions in a skyrmion crystal work as batteries in series and so can provide a large electric voltage by summing each contribution. The simulations give electric voltages of a few microvolts when using a nanoscopically small system of $\sim$50 nm $\times$ 50 nm containing twelve skyrmions. This indicates that if we use a larger device or sample with a microscopic size rather than a nanoscopic sample, the generated electric voltage should be five or six orders of magnitude greater than that obtained in the present simulations. These findings pave a way to efficient conversion of microwaves to DC electric voltage via using skyrmion-hosting magnets, which should be useful in future spintronics devices. Experimental realization and observation of the phenomena predicted here are subjects for future research.

\section{Acknowledgments}
This work was partly supported by JSPS KAKENHI (Grants No. 17H02924, No. 16H06345, and No. 19H00864) and Waseda University Grant for Special Research Projects (Project No. 2019C-253). We thank J. Ohe, M. Ikka, Y. Ohki, and Y. Shimada for useful discussions.


\begin{thebibliography}{999}
\bibitem{Volovik87}G. E. Volovik, J. Phys. C {\bf 20}, L83 (1987).

\bibitem{Barnes07}S. E. Barnes and S. Maekawa, Phys. Rev. Lett. {\bf 98}, 246601 (2007).

\bibitem{SAYang09}S. A. Yang, G. S. D. Beach, C. Knutson, D. Xiao, Q. Niu, M. Tsoi, and J. L. Erskine, Phys. Rev. Lett. {\bf 102}, 067201 (2009).

\bibitem{Tanabe12}K. Tanabe, D. Chiba, J. Ohe, S. Kasai, H. Kohno, S. E. Barnes, S. Maekawa, K. Kobayashi, and T. Ono, Nat. Comm. {\bf 3}, 845 (2012).
\bibitem{Mochizuki12}M. Mochizuki, Phys. Rev. Lett. {\bf 108}, 017601 (2012).

\bibitem{Petrova11}O. Petrova and O. Tchernyshyov, Phys. Rev. B {\bf 84}, 214433 (2011).
\bibitem{Ohe13}J. Ohe and Y. Shimada, Appl. Phys. Lett. {\bf 103}, 242403 (2013).

\bibitem{Shimada15}Y. Shimada and J. I. Ohe, Phys. Rev. B {\bf 91}, 174437 (2015).
\bibitem{Bak80}P. Bak, and M. H. Jensen, J. Phys. C {\bf 13}, L881 (1980).

\bibitem{YiSD09}S. D. Yi, S. Onoda, N. Nagaosa, and J. H. Han, Phys. Rev. B {\bf 80}, 054416 (2009).
\bibitem{Bogdanov89}A. N. Bogdanov and D.A. Yablonskii, Sov. Phys. JETP {\bf 68}, 101 (1989).

\bibitem{Bogdanov94}A. Bogdanov and A. Hubert, J. Mag. Mag. Mat. {\bf 138}, 255 (1994).

\bibitem{Rossler06}U.K. R\"o{\ss}ler, A. N. Bogdanov, and C. Pfleiderer, Nature {\bf 442}, 797 (2006).

\bibitem{Nagaosa13}N. Nagaosa and Y. Tokura, Nat. Nanotech. {\bf 8}, 899 (2013).
\bibitem{Fert13}A. Fert, V. Cros, and J. Sampaio, Nat. Nanotech. {\bf 8}, 152 (2013).

\bibitem{Mochizuki15a}M. Mochizuki and S. Seki, J. Phys.: Cond. Matt. {\bf 27}, 503001 (2015).

\bibitem{Seki15}S. Seki and M. Mochizuki, ``Skyrmions in Magnetic Materials" (Springer Briefs in Physics).

\bibitem{Muhlbauer09}S. M\"uhlbauer, B. Binz, F. Jonietz, C. Pfleiderer, A. Rosch, A. Neubauer, R. Georgii, and P. B\"oni, Science {\bf 323}, 915 (2009).

\bibitem{YuXZ10}X. Z. Yu, Y. Onose, N. Kanazawa, J. H. Park, J. H. Han, Y. Matsui, N. Nagaosa, and Y. Tokura, Nature {\bf 465}, 901 (2010).

\bibitem{Seki12}S. Seki, X. Z. Yu, S. Ishiwata, and Y. Tokura, Science {\bf 336}, 198 (2012).

\bibitem{Adams12}T. Adams, A. Chacon, M. Wagner, A. Bauer, G. Brandl, B. Pedersen, H. Berger, P. Lemmens, and C. Pfleiderer, Phys. Rev. Lett. {\bf 108}, 237204 (2012).

\bibitem{Tonomura12}A. Tonomura X. Z. Yu, K. Yanagisawa, T. Matsuda, Y. Onose, N. Kanazawa, H. S. Park, and Y. Tokura, Nano Lett. {\bf 12}, 1673 (2012).

\bibitem{LinSZ14}S.-Z. Lin, C. D. Batista, and A. Saxena, Phys. Rev. B {\bf 89}, 024415 (2014).

\bibitem{Schwarze15}T. Schwarze, J. Waizner, M. Garst, A. Bauer, I. Stasinopoulos, H. Berger, C. Pfleiderer, and D. Grundler, Nat. Mater. {\bf 14}, 478 (2015).

\bibitem{Garst17}M. Garst, J. Waizner, and D. Grundler, J. Phys. D: Appl. Phys. {\bf 50}, 293002 (2017).

\bibitem{Finocchio16}G. Finocchio, F. B\"uttner, R. Tomasello, M. Carpentieri, and M. Kl\"aui, J. Phys. D: Appl. Phys. {\bf 49}, 423001 (2016).
\bibitem{Mochizuki13}M. Mochizuki, and S. Seki, Phys. Rev. B {\bf 87}, 134403 (2013).

\bibitem{Okamura13}Y. Okamura, F. Kagawa, M. Mochizuki, M. Kubota, S. Seki, S. Ishiwata, M. Kawasaki, Y. Onose, Y. Tokura, Nat. Commun. {\bf 4}, 2391 (2013).

\bibitem{Mochizuki15b}M. Mochizuki, Phys. Rev. Lett. {\bf 114}, 197203 (2015).

\bibitem{Okamura15}Y. Okamura, F. Kagawa, S. Seki, M. Kubota, M. Kawasaki, and Y. Tokura, Phys. Rev. Lett. {\bf 114}, 197202 (2015).

\bibitem{YuXZ11}X. Z. Yu, N. Kanazawa, Y. Onose, K. Kimoto, W. Z. Zhang, S. Ishiwata, Y. Matsui, and Y. Tokura, Nat. Mater. {\bf 10} 106 (2011).

\bibitem{LinSZ15}S.-Z. Lin and A. Saxena, Phys. Rev. B {\bf 92}, 180401(R) (2015).
\bibitem{Karube16}K. Karube, J. S. White, N. Reynolds, J. L. Gavilano, H. Oike, A. Kikkawa, F. Kagawa, Y. Tokunaga, H. M. Ronnow, Y. Tokura and Y. Taguchi, Nat. Mater. {\bf 15}, 1237 (2016).

\bibitem{Nakajima17}T. Nakajima, H. Oike, A. Kikkawa, E. P. Gilbert, N. Booth, K. Kakurai, Y. Taguchi, Y. Tokura, F. Kagawa and T. Arima, Sci. Adv. {\bf 3}, e1602562 (2017).

\bibitem{Oike16}H. Oike, A. Kikkawa, N. Kanazawa, Y. Taguchi, M. Kawasaki, Y. Tokura and F. Kagawa, Nat. Phys. {\bf 12}, 62 (2016).

\bibitem{Chacon18}A. Chacon, L. Heinen, M. Halder, A. Bauer, W. Simeth, S. Muhlbauer, H. Berger, M. Garst, A. Rosch, and C. Pfleiderer, Nat. Phys. {\bf 14}, 936 (2018).
\bibitem{Matsumoto16}T. Matsumoto, Y.-G. So, Y. Kohno, H. Sawada, Y. Ikuhara and N. Shibata, Sci. Adv. {\bf 2}, e1501280 (2016).

\bibitem{Nakajima17b}H. Nakajima, A. Kotani, M. Mochizuki, K. Harada and S. Mori, Appl. Phys. Lett. {\bf 111}, 192401 (2017).
\bibitem{WangW15}W. Wang, M. Beg, B. Zhang, W. Kuch, and H. Fangohr, Phys. Rev. B {\bf 92}, 020403(R) (2015).

\bibitem{Ikka18}M. Ikka, A. Takeuchi, and M. Mochizuki, Phys. Rev. B 98, 184428 (2018).
\end{thebibliography}
\end{document}